\documentclass[a4paper,fleqn,usenatbib]{mnras}

\usepackage{newtxtext,newtxmath}
\usepackage[T1]{fontenc}
\usepackage{ae,aecompl}
\usepackage[english]{babel}
\usepackage{graphicx}	% Including figure files
\usepackage{subfigure}
\usepackage{amsmath}	% Advanced maths commands
\usepackage{amssymb}	% Extra maths symbols

\title[New cLBV in NGC\,4736]{New Luminous Blue Variable Candidates in NGC\,4736}

\author[Y. Solovyeva et al.]{
Y. Solovyeva,$^{1}$\thanks{E-mail: solovyeva@sao.ru}
A. Vinokurov,$^{1}$
S. Fabrika,$^{1,2}$
A. Kostenkov,$^{1,3}$
O. Sholukhova,$^{1}$
\newauthor
A. Sarkisyan,$^{1}$
A. Valeev,$^{1,2}$
 K. Atapin,$^{1,4}$
 O. Spiridonova,$^{1}$
 A. Moskvitin,$^{1}$
 E. Nikolaeva$^{2}$
\\
$^{1}$Special Astrophysical Observatory, Nizhnij Arkhyz, Russia\\
$^{2}$Kazan Federal University, Kremlevskaya 18, 420008 Kazan, Russia\\
$^{3}$Saint Petersburg State University, 7/9 Universitetskaya Emb., 199034, Saint Petersburg, Russia\\
$^{4}$ Sternberg Astronomical Institute, Moscow State University, Universitetskij Pr. 13, Moscow 119992, Russia
}

\date{Accepted XXX. Received YYY; in original form ZZZ}

\pubyear{2018}

\begin{document} \label{firstpage} \pagerange{\pageref{firstpage}--\pageref{lastpage}} \maketitle

\begin{abstract} 
 
We have found three new LBV candidates in the star-forming galaxy NGC\,4736. They show typical well-known LBV spectra, broad and strong hydrogen lines, \ion{He}{i} lines, many \ion{Fe}{ii} lines, and forbidden [\ion{Fe}{ii}] and [\ion{Fe}{iii}]. Using archival Hubble Space Telescope and ground-based telescope data, we have estimated the bolometric magnitudes of these objects from -8.4 to -11.5, temperatures, and reddening. Source NGC\,4736\_1 ($M_V =-10.2 \pm 0.1 $ mag) demonstrated variability between 2005 and 2018 as $\Delta V \approx 1.1$ mag and $\Delta B \approx 0.82$ mag, the object belongs to LBV stars. 
 NGC\,4736\_2 ($M_V < -8.6$ mag) shows P\,Cyg profiles and its spectrum has changed from 2015 to 2018. The brightness variability of NGC\,4736\_2 is $\Delta V \approx 0.5$ mag and $\Delta B \approx 0.4$ mag. In NGC\,4736\_3 ($M_V=-8.2\pm 0.2$ mag), we found strong nebular lines, broad wings of hydrogen; the brightness variation is only $\approx 0.2$ mag. Therefore, the last two objects may reside to LBV candidates. 

\end{abstract} 


%%%%%%%%%%%%%%%%%%%%%%%%%%%%%%%%%%%%%%%%%%%%%%%%%%
%%%%%%%%%%%%%%%%% BODY OF PAPER %%%%%%%%%%%%%%%%%%
\section{Introduction} 

Luminous blue variables (LBVs) are bright massive stars at one of their final evolutionary phases \citep{Humphreys1994}. This evolutionary phase is characterized by a high mass-loss rate due to a stellar wind and also by matter ejection 
in a strong brightness. 
Many LBVs are surrounded by compact circumstellar envelopes of different morphology \citep{Nota1995,Weis2001} that originate during the brightness increase. 

LBV stars show a strong spectral and photometric variability of  different amplitudes at  time scales  of months and years \citep{vanGenderen2001}.
When star brightness in the visual range reaches its maximum, the photosphere temperature
falls down to 7000--8000~K (cold state). 
Here the observed spectrum is similar to those of F stars \citep{Massey2007}.  
With a decrease of the apparent brightness, the temperature of a star can reach more than 35000 K \citep{Clark2005} (hot state).
In this state, the LBV have WNLh-star-like spectra. 

It is known that massive O stars with the masses $M > 40 M_{\odot}$ \citep{Maeder1996} can pass the LBV phase, after which they enter the late WR-star phase of the nitrogen sequence (WNL).There are less luminous LBVs with initial masses $\sim 25-40 M_{\odot}$ that can become red supergiants \citep{Humphreys2016}.
However, in many studies of the last decade \citep{Galyam2009, Kotak2006, Groh2013}, it is supposed that LBVs can be immediate supernovae progenitors.

Currently, only a small number of LBV stars are known which makes objects of this type unique. 
The search for LBVs in our Galaxy is a difficult task due to high extinction in the galactic plane and not good the distance measurements. However, modern IR studies have made it possible to increase the number of known LBV stars and similar objects in our Galaxy \citep{Clark2003,Gvaramadze2010a,Gvaramadze2010b}. The search for LBV-like objects in nearby galaxies is also of great interest, since the distances to them are determined reliably. 

Initially, the LBV stars were searched for by their photometric and spectral variability in optics \citep{Hubble1953}.
It took tens of years to detect the variability, so other methods have been proposed.
For example, searching for SS433-like objects from spectroscopic data \citep{Fabrika1995,Calzetti1995} which are really similar to the spectra of LBV stars in a hot state \citep{Fabrika2015}, and also conducting the UV observations \citep{Massey1996}. The basic method is searching from the H$\alpha$ emissions associated with blue stars \citep{Valeev2009,Sholukh1997, Valeev2010}. 
Besides LBV stars, there are a lot of similar objects such as B[e] supergiants \citep{Fabrika2005,Sholukh2015}, \ion{Fe}{ii} emission line stars, warm supergiants, hot supergiants, yellow supergiants \citep{Humphreys2017}. 

In this work, we have also looked at the presence of H$\alpha$ emissions in blue point-like star objects. To search for objects of interest, we have chosen the NGC~4736 galaxy with face-on orientation and the distance of $m-M = 28.31 \pm 0.08$ \citep{Tully2013}.

\section{Target selection} 
\begin{figure*} \vspace{-4ex} 
\centering 
\subfigure[]{ 
\includegraphics[width=0.21\linewidth]{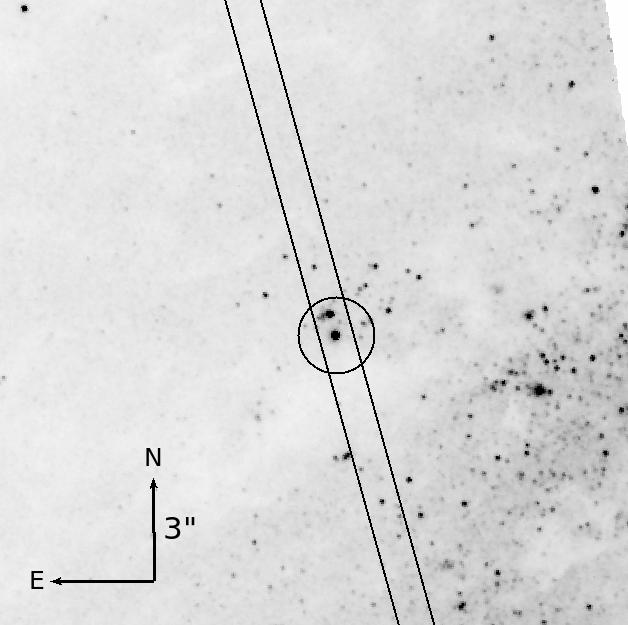} 
\label{Fig:1a} }
\hspace{4ex} 
\subfigure[]{ 
\includegraphics[width=0.202\linewidth]{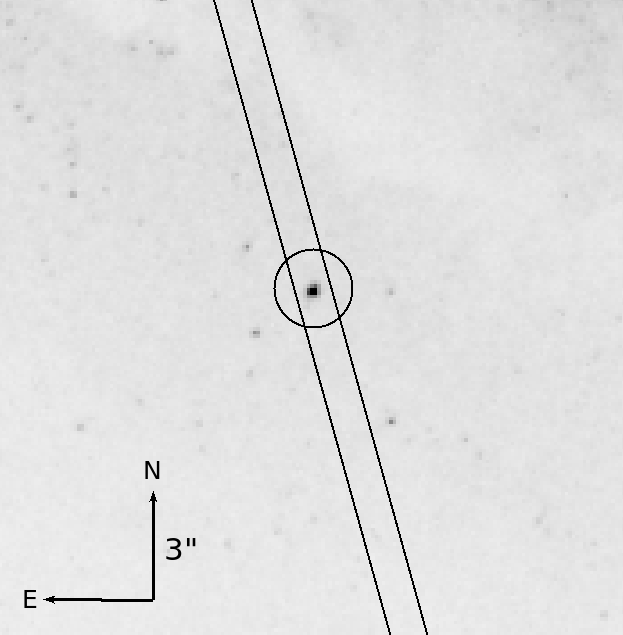} 
\label{Fig:1b} } 
\hspace{4ex} 
\subfigure[]{ 
\includegraphics[width=0.21\linewidth]{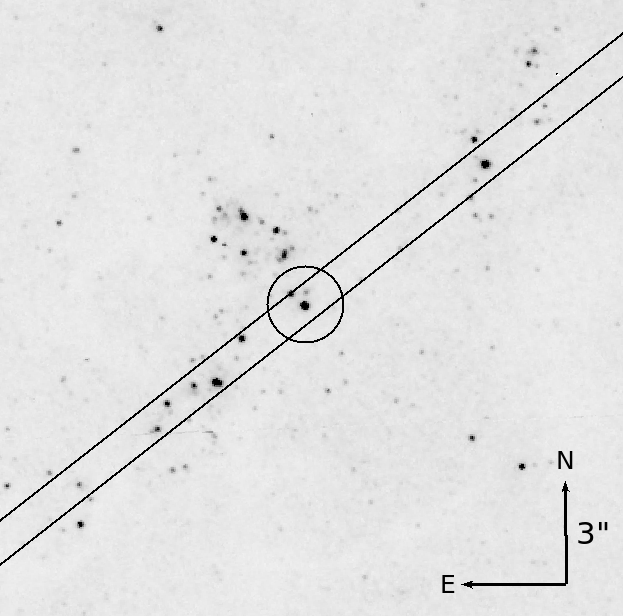} 
\label{Fig:1c} } 
\caption{Images in HST/F555W: (a) NGC\,4736\_1, (b) NGC\,4736\_2, and (c) NGC\,4736\_3. The circles mark the central objects
(a) with the radius of 1.1\arcsec, it is the maximum seeing on the ground-based images, the 1" slit is shown.} 
\label{Fig1} 
\end{figure*}

For preliminary search we used archival HST images obtained with ACS, WFPC2, and WFC3 cameras with broadband and narrowband (F656N, F657N and F658N) filters. The criterion was: a source should be point-like and bright in all the filters.
We have found three LBV candidates (Fig.\ref{Fig1}). Their HST coordinates are 12:50:57.264, +41:07:23.13 (NGC\,4736\_1), 12:50:55.844, +41:06:25.44 (NGC\,4736\_2) and  12:51:03.358, +41:06:35.37 (NGC\,4736\_3).

\section{Observations and results} 
We analysed the spectral and photometric variability of these three sources. 
The optical spectra were obtained with the 6-m BTA telescope in 2014-2018 using the SCORPIO optical reducer within the range of 4000-5700 \AA\ and 5700-7500 \AA, the resolution was 5.3 \AA. 
Object NGC\,4736\_1 was observed on January 3, 2014 (VPHG1200R) and on January 18, 2015 (VPHG1200G). Object NGC\,4736\_2  was observed on January 3, 2014 (VPHG1200R), January 18, 2015 (VPHG1200G) and February 18, 2018 (VPHG1200G).
The spectra of NGC\,4736\_3 were taken on January 3, 2014 (VPHG1200R) and on March 31, 2017. 
In all three targets the seeing was from 1.2\arcsec\, to 1.6\arcsec.
Spectroscopic data reduction was carried out with the LONG context in the MIDAS using standard algorithm. 
An extraction of the spectra have been done with the SPEXTRA package \citep{Sarkis2017} developed for long-slit 2D spectra in crowded stellar fields. 

The photometric data have been obtained with the 1-m Zeiss-1000 and 6-m BTA telescope of Special Astrophysical Observatory of RAS and also with the 2.5-m telescope of the Caucasian Mountain Observatory of SAI MSU. Moreover, we have used archival data of the ground-based telescopes (KPNO 2.1-m, JKT, Palomar 60-inch) and the HST data. Observation dates, instruments, filters and magnitudes are given in Table \ref{obsphot}. We conducted the primary photometric data reduction in the MIDAS environment. Measurements of apparent magnitudes were done by the aperture photometry method using 11-22 selected reference stars with the APPHOT package in the IRAF. To reduce the obtained magnitudes into the standard Johnson-Cousins system, the PySynphot package was used.

\begin{figure} 

\includegraphics[angle=0,scale=0.47]{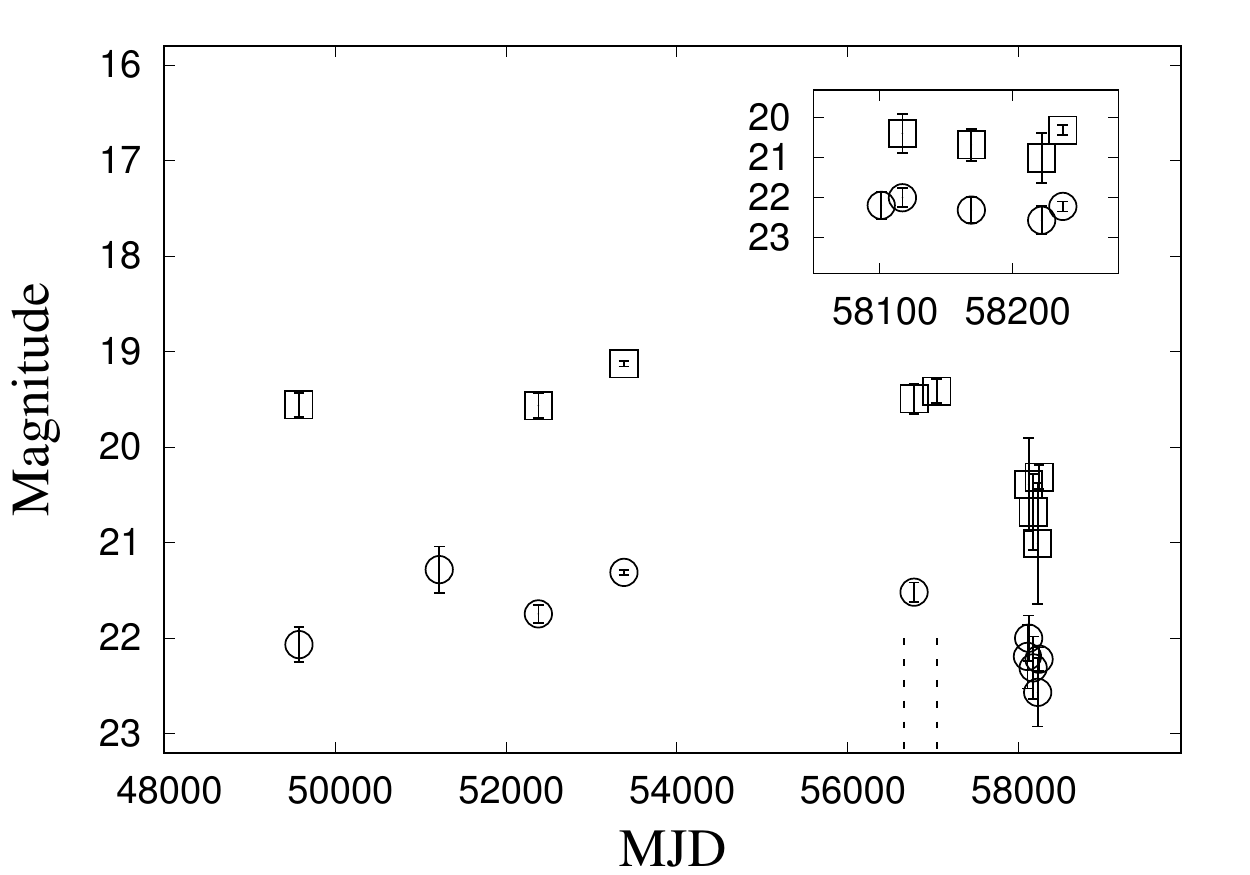} 
\includegraphics[angle=270,scale=0.225]{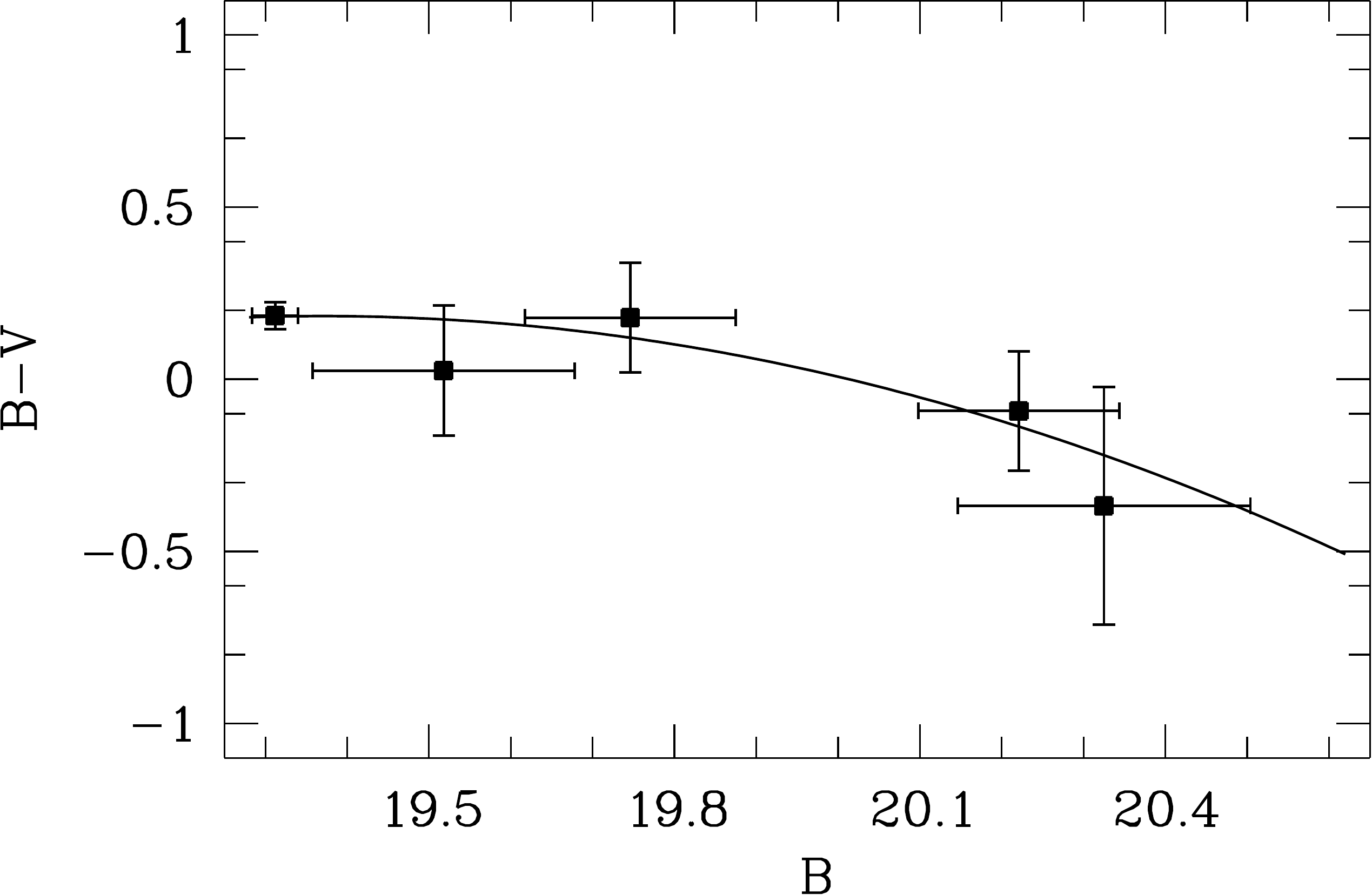} 
\includegraphics[angle=270,scale=0.225]{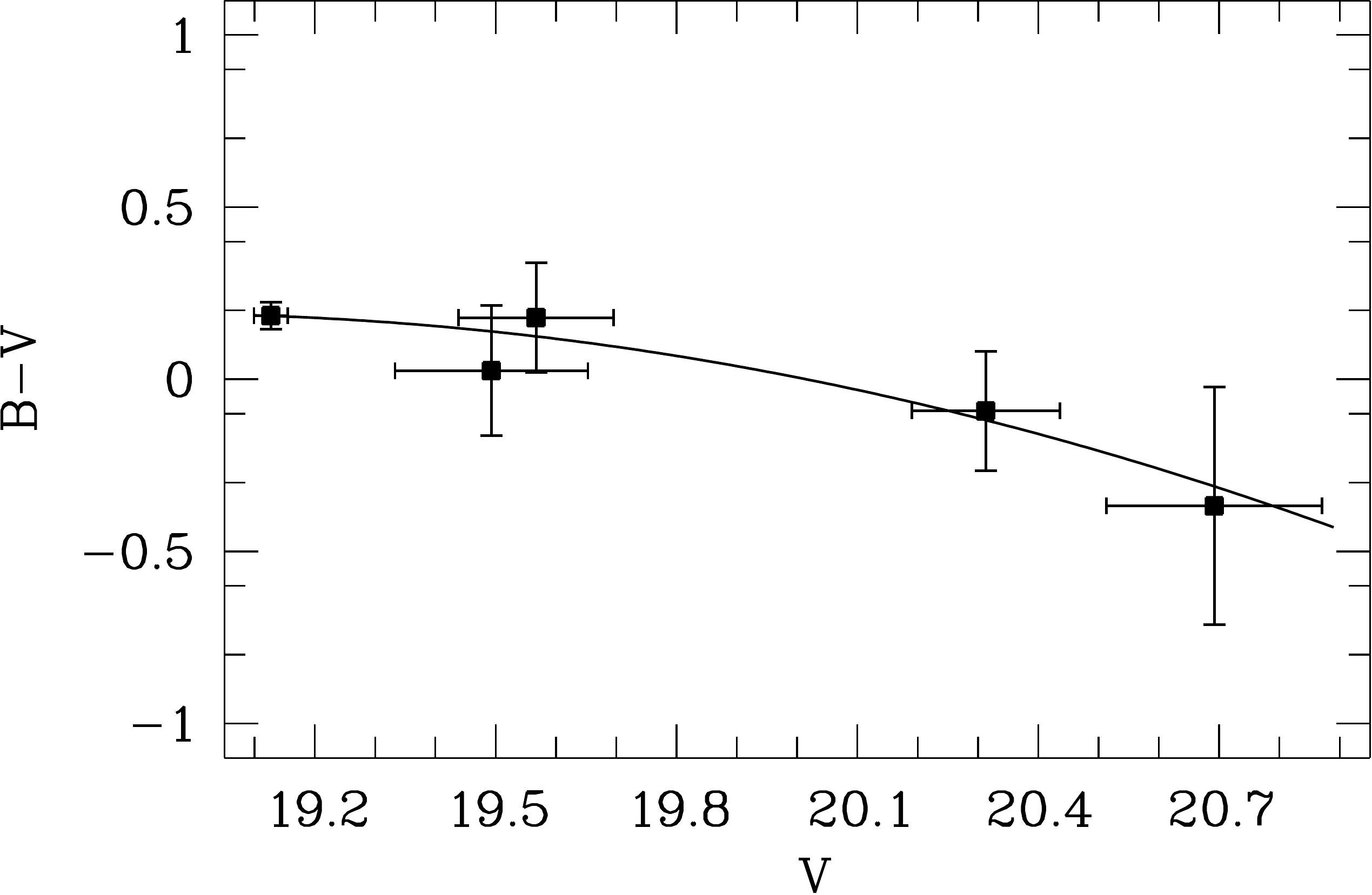} 
\caption{(top) Light curve of NGC\,4736\_1 in the B (circles) and V (squares) bands. The B band is shifted down by two magnitudes. The insert expands a region between MJD 58100--58250. The photometric data are shown in Table\,\ref{obsphot}. Two dashed lines represent dates of the spectroscopic observations, the blue spectrum is on the right (January 18, 2015) and the red spectrum is on the left (January 3, 2014). (middle) The colour versus the B magnitude and (bottom) versus the V magnitude. The solid line denotes the quadratic polynomials.} 
\label{Fig2} 
\end{figure}

\begin{table*}
\caption{Log of the photometric data. The columns show the instruments, dates and magnitudes in Johnson-Cousins system. In the last three columns magnitudes of NGC\,4736\_1, NGC\,4736\_2, NGC\,4736\_3 in B and V bands are given slash separated. } 
\begin{tabular}{lccccc} \hline\hline  \centering
Telescope & Date & Filter & \multicolumn{2}{|c|}{\quad\quad\quad\quad\quad B/V magnitude} \\ 
 & & \multicolumn{4}{|c|}{\quad\quad\quad\quad\quad\quad\quad\quad NGC\,4736\_1\quad\quad\quad\quad\quad\quad\quad  NGC\,4736\_2\quad\quad\quad\quad\quad\quad\quad  NGC\,4736\_3}\\ \hline\hline
JKT & 1994-08-12  & Harris$^a$ & $20.07\pm0.19$/$19.56\pm0.12$ & $20.05\pm0.11$/$19.95\pm0.10$ & --/-- \\ \hline
Palomar 60 inch & 1999-02-08 & Johnson-Cousins & $19.28\pm0.24$/-- & --/-- & --/-- \\ \hline
KPNO 2.1m & 2002-04-14 & Harris$^a$ & $19.75\pm0.09$/$19.57\pm0.13$ & $20.05\pm0.05$/$19.99\pm0.07$ & $20.78\pm0.05$/$20.66\pm0.07$ \\ \hline
HST& 2005-01-10 & HST filter system$^b$ & $19.31\pm0.03$/$19.13\pm0.03$ & $19.80\pm0.03$/$19.72\pm0.03$ & $20.77\pm0.03$/$20.58\pm0.03$\\ \hline
BTA & 2014-04-27 &Johnson-Cousins & $19.52\pm0.10$/$19.49\pm0.16$ & $19.77\pm0.05$/$19.66\pm0.06$ & --/-- \\ \hline
BTA & 2015-01-18 & Johnson-Cousins &--/$19.41\pm0.13$ & --/$20.05\pm0.07$ & --/$20.77\pm0.09$\\ \hline
BTA & 2017-03-31 & Johnson-Cousins &--/$20.42\pm0.42$ & --/$19.92\pm0.10$ & --/$20.51\pm0.09$\\ \hline
Zeiss1000 & 2017-12-13 & Johnson-Cousins & $20.19\pm0.33$/-- & $19.76\pm0.08$/-- & --/--\\ \hline
Zeiss1000  & 2017-12-29 & Johnson-Cousins & $20.00\pm0.24$/$20.39\pm0.49$ & $19.83\pm0.08$/$19.72\pm0.08$& --/-- \\ \hline
BTA  & 2018-02-19 & Johnson-Cousins & $20.31\pm0.33$/$20.68\pm0.4$ & $20.01\pm0.10$/$20.02\pm0.11$ & --/--\\ \hline
BTA  & 2018-04-13 & Johnson-Cousins & $20.57\pm0.36$/$21.01\pm0.63$ & $20.04\pm0.08$/$19.99\pm0.09$ & --/--\\ \hline
2.5-m,SAI MSU  & 2018-04-29 & Johnson-Cousins & $20.22\pm0.12$/$20.31\pm0.12$ & $20.20\pm0.05$/$20.21\pm0.06$ & $20.83\pm0.11$/$20.71\pm0.07$\\ \hline
\end{tabular} 
\label{obsphot}
\begin{center}
\textit{Notes.}  
$^a$The photometry obtained in Harris system and converted to Johnson-Cousins system. $^b$The data were obtained in filters F435W and F555W of HST and converted to Johnson-Cousins system.
\end{center}
\end{table*}

\subsection{NGC\,4736\_1} 

\label{sec:obj1} 

\begin{figure*} 
\begin{center} 
\includegraphics[angle=270,scale=0.41]{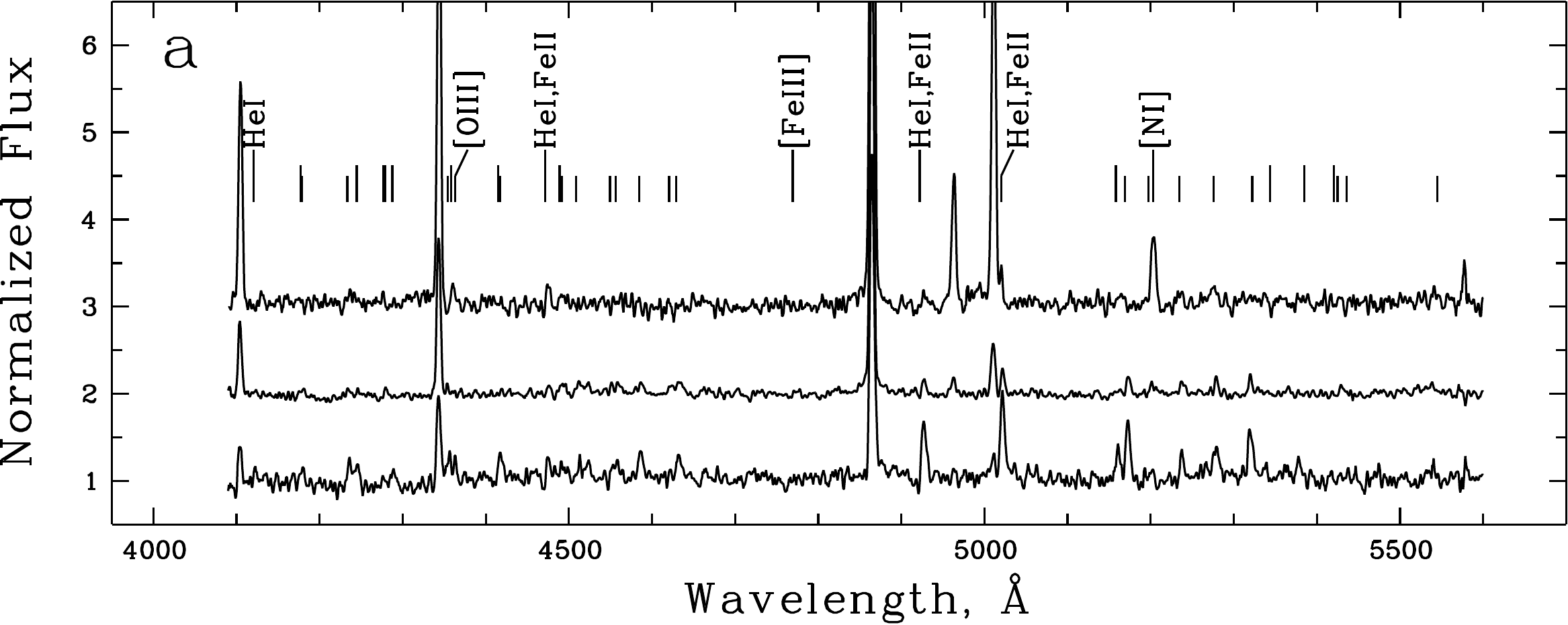} 
\includegraphics[angle=270,scale=0.41]{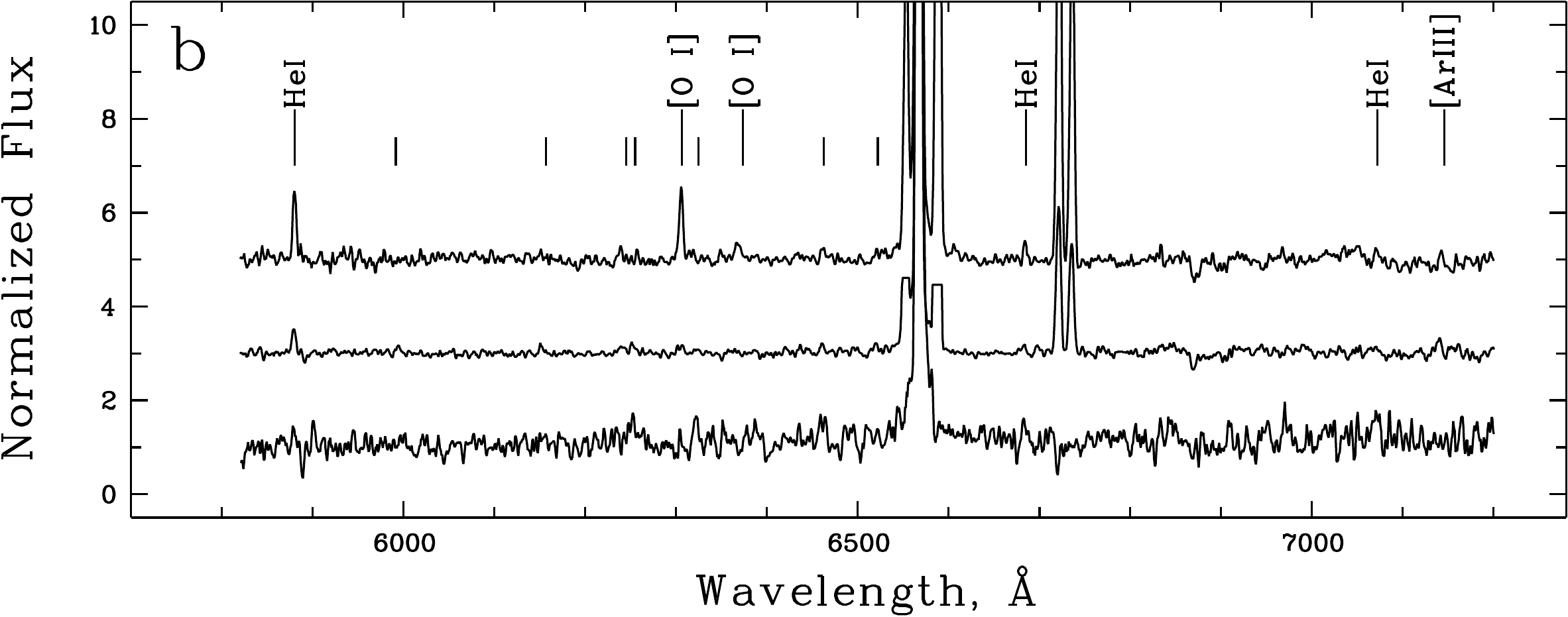} 
\caption{Blue (a) and red (b) spectra of NGC\,4736\_3 (top), NGC\,4736\_1 (middle) and NGC\,4736\_2 (bottom). The unlabelled short and long ticks designate the \ion{Fe}{ii} and [\ion{Fe}{ii}] lines, respectively. Narrow components of H$\alpha$, H$\beta$, H$\gamma$ and H$\delta$, the forbidden lines [\ion{O}{iii}], [\ion{N}{i}], [\ion{O}{i}], [\ion{N}{ii}] and [\ion{S}{ii}] belong to the surrounding nebulae. We have removed [\ion{N}{ii}] lines  from middle red spectrum.} 
\label{Fig3} 
\end{center} 
\end{figure*}

The object NGC\,4736\_1 is located in the crowded field. The bright star is located about 0.64\arcsec\, from the target (Fig.\ref{Fig1}\,a) and its brightness is about 1 mag less than NGC\,4736\_1. Other stars are significantly weaker. The ground-based telescopes cannot resolve them. Therefore, for each ground-based observation, we measured the total flux of all the stars falling into the given aperture (aperture radii were from 0.75\arcsec\, to 1.3\arcsec\, depending on the seeing in a particular observation) and then separated out contribution of the target and other stars within this aperture using HST images. To do this, we obtained both the total flux and the target flux (by measuring in small 3-pixel aperture and applying aperture corrections) in the same aperture of HST images and calculated the contribution of other stars. 
Using this contribution and total fluxes and assuming the constancy of the field stars, 
we found the target flux in the ground-based images. 

The total flux errors are in the range $0.07 - 0.16$ mag (the apertures 0.75\arcsec\, -- 1.3\arcsec\,), and after excluding the contribution of the field stars the target magnitudes and errors are in Table~\ref{obsphot}, and in Fig.~\ref{Fig2}.

Despite a low accuracy of the magnitude determination (Fig.~\ref{Fig2}, top), the brightness of NGC\,4736\_1 shows clear downtrend. Variations of the brightness reach $\Delta$V $\approx 1.1$ mag and $\Delta$B $\approx 0.8$ 2mag in the V and B bands respectively. It is obvious that such a strong brightness changes, and can not be explained either by the calibration errors or by the errors associated with stellar flux estimation. 
 
In Fig.~\ref{Fig2} we show the B--V colour versus the brightness variation in the B and V bands and its approximation by a quadratic polynomial. The dependence is $ B-V = -0.442(B-19.0)^2 + 0.326(B-19.0) + 0.123$ (Fig.~\ref{Fig2}, middle) and 
$ B-V = -0.161(V-19.0)^2 - 0.023(V-19.0) + 0.190$ (Fig.\ref{Fig2}, bottom). 
That yields ${\chi^2}/dof$ = 0.48 and ${\chi^2}/dof$ = 0.26 with degree of freedom  $dof=2$.
This kind of dependence is characteristic of the well-known LBV star V532 (Romano star) in M33 (\cite{Sholukh2011}, Fig.3). This supports the fact that the temperature of the stellar photosphere increases with its brightness decrease. 

The observed \ion{He}{i} lines are pronounced with the temperature about 20 kK (and decrease to 15 kK and 25 kK). The \ion{Fe}{ii} lines have a maximum intensity when temperature is about 15 kK, and they decrease to 10 and 20 kK. Therefore, we have estimated the photosphere temperature of NGC\,4736\_1 T = 18 $\pm$ 3 kK. Based on the intrinsic colour $(B-V)_0$ for supergiants \citep{Fitzgerald1970,Straizys1981} we have estimated the photosphere temperature from HST data as $17 \pm 3$ kK (dereddened colour $(B-V)_0 = -0.14 \pm 0.04$). This estimate is consistent with temperature from spectral lines.

In addition to broad components of the H$\alpha$, H$\beta$, H$\gamma$, and H$\delta$ lines, in the NGC\,4736\_1 (Fig.~\ref{Fig3}, middle spectra), there are many bright emission lines of \ion{Fe}{ii}. The \ion{He}{i} lines and forbidden lines [\ion{Fe}{ii}], [\ion{Fe}{iii}] are emitted by the surrounding wind, however, the lines [\ion{Ar}{iii}], [\ion{N}{ii}], [\ion{S}{ii}] may belong to the nebula.
 
From the hydrogen lines of the surrounding nebula, assuming the B case photoionization  \citep{Oster2006}, we found the reddening  $A_V = 1.0 \pm 0.1$ mag. From this we found the absolute magnitude $M_V=-10.2 \pm 0.1$ mag and the bolometric magnitude $M_{bol}= - 11.5 \pm 0.5$ mag. The bolometric luminosity of the candidate NGC\,4736\_1 is about $log(L_{bol}/L_{\odot}) \approx 6.5 \pm 0.2$, that is comparable to those of $\eta$ Car \citep{Cox1995}. The NGC\,4736\_1 spectrum, its bolometric luminosity, and also a considerable brightness variability ($\Delta$V $\approx1.1$ mag and $\Delta$B $\approx 0.82$ mag) allow us to claim that NGC\,4736\_1 is the LBV star. 
\subsection{NGC\,4736\_2} 
\label{sec:obj2} 

\begin{figure*} 
\centering
\subfigure[]{
\includegraphics[angle=270,scale=0.41]{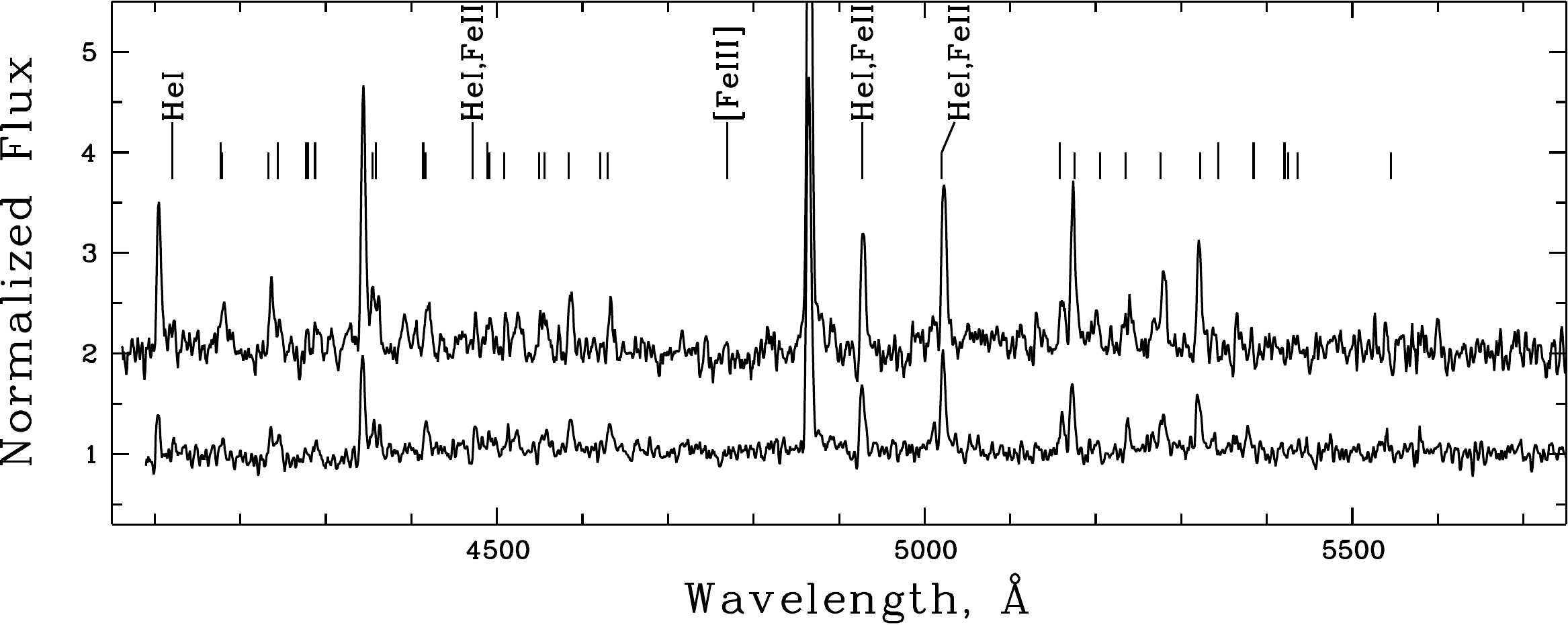} 
\label{Fig:4a}}
\subfigure[]{
\includegraphics[angle=270,scale=0.26]{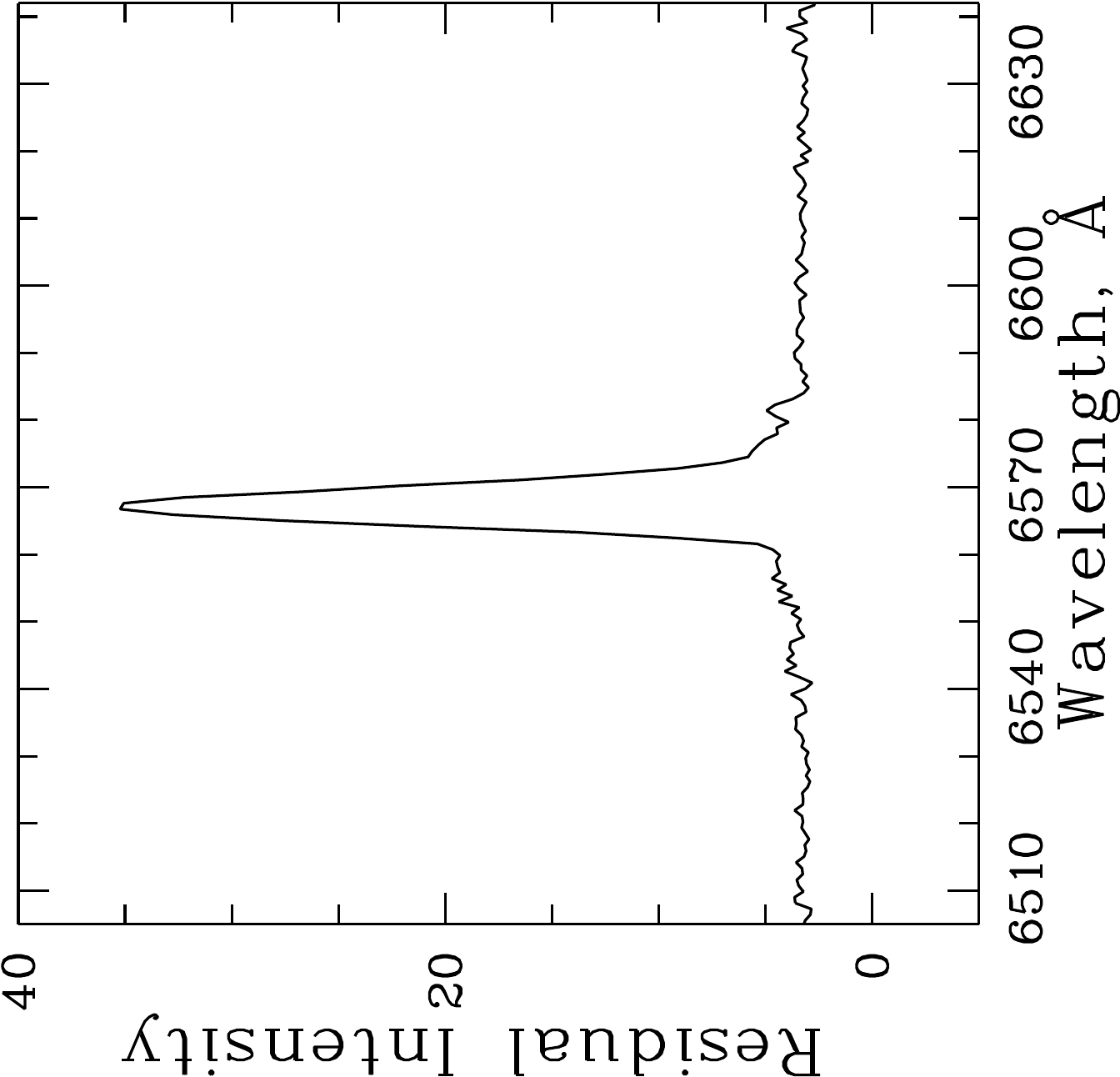}
\label{Fig:4b}}
\caption{
(a)Blue spectra of NGC\,4736\_2 taken in 2015 (bottom) and in 2018 (top). The unlabelled short and long ticks designate the \ion{Fe}{ii} and [\ion{Fe}{ii}] lines, respectively.(b)   H$\alpha$ with the hidden P Cyg profile taken in 2014.
} 
\label{Fig4} 
\end{figure*}

 In the blue spectrum (Fig.~\ref{Fig3}\,a, bottom) we observe \ion{He}{i} lines with P Cyg profiles and numerous \ion{Fe}{ii} lines. H$\beta$ and H$\alpha$ lines (Fig. \ref{Fig:4b}) demonstrate the hidden P Cyg profiles. A large number of forbidden iron lines [\ion{Fe}{ii}], [\ion{Fe}{iii}] indicate the outflow of the stellar matter as a powerful wind.

The \ion{He}{i} lines (4921, 5015) also show the P Cyg profile. In the spectrum obtained in 2018 (Fig. \ref{Fig4}), the spectral line variability was found, the Balmer lines noticeably broadened and the blue wing of the lines became less evident. Moreover, the brightness decrease in the V band was $\Delta$V $\approx 0.2$ mag during the period from 2015 to 2018. 

Generally, NGC\,4736\_2 blue spectrum is similar to this of NGC\,4736\_1, but \ion{Fe}{ii} lines are stronger than the same lines in NGC\,4736\_1. Therefore, we estimated NGC\,4736\_2 photosphere temperature as $15 \pm 3 kK$.
There is no nebula associated with NGC\,4736\_2, so we could not estimate the reddening Av. Therefore we estimated only lower limits of the absolute and bolometric magnitudes as $M_V<-8.6$ mag and $M_{bol}<-10.0$ mag, respectively. With given estimates of temperature of the photosphere, we obtained estimate of the bolometric luminosity of NGC\,4736\_2 $log(L_{bol}/L_{\odot})> 5.9$.

Using the data obtained with the 6-m BTA telescope in 2014 and 2018, the brightness variations of the candidate NGC\,4736\_2 (Fig. \ref{Fig1}\,b) can be estimated as $\Delta$B $\approx 0.4$ mag in the B band and $\Delta$V $\approx 0.5$ mag in the V band. Such brightness variations are not enough to classify NGC 4736\_2 as an LBV, so it remains in the status of the candidate. 

\subsection{\textbf{NGC\,4736\_3}} 
\label{sec:obj3} 
 
NGC\,4736\_3 shows the narrow Balmer lines, \ion{He}{I} lines, forbidden [\ion{O}{iii}], [\ion{O}{i}], [\ion{N}{i}], [\ion{N}{ii}], and [\ion{S}{ii}] lines. The broad components of Balmer lines are observed as a wind. Due to the presence of broad H$\alpha$, H$\beta$ and weak \ion{Fe}{ii} lines we have estimated temperature as $12 \pm 2$ kK. Using HST dereddened colour $(B-V)_0 = 0.04 \pm 0.04$ we obtained lower temperature $9.3 \pm 1.0$ kK.

Using the flux ratios of the hydrogen nebular lines $H\delta/H\beta$ and $H\gamma/ H\beta$, we estimated the reddening $A_V = 0.47 \pm 0.21$ mag. We determined the absolute and bolometric magnitudes as $M_V=-8.2 \pm 0.2$ mag and $M_{bol} = -8.4 \pm 0.4$ mag, respectively. The obtained estimates of the photosphere temperature and reddening made it possible to estimate the bolometric luminosity of the object $log(L_{bol}/L_{\odot})\approx 5.3 \pm 0.2$.

Brightness variability of the candidate NGC\,4736\_3 (Fig. \ref{Fig1}\,c)
is inconsiderable and reaches $0.2$ mag in the B and V filters. However, we need a better spectrum and photometry, when the variability will appear. The candidate NGC\,4736\_3 is LBV-like according to both its spectrum and bolometric luminosity.
However, the insufficiency of photometric data leaves NGC\,4736\_3 in the status of 
the LBV candidate. 
 
\section{Discussion and conclusions} 

All the LBV candidates that we have studied, have the spectra similar to the spectra of the known LBVs. This is  evident from the presence of broad hydrogen lines,\,
\ion{He}{i} lines, numerous iron \ion{Fe}{ii}, [\ion{Fe}{ii}], [\ion{Fe}{iii}] emission lines. Their bolometric luminosities are also typical of stars of this type, the luminosities of NGC\,4736\_1, NGC\,4736\_2, and NGC\,4736\_3 are of the order of $log(L_{bol}/L_{\odot}) \approx 6.5 \pm 0.2$, $log(L_{bol}/L_{\odot}) > 5.9$, and $log(L_{bol}/L_{\odot}) \approx 5.3 \pm 0.2$, respectively.

The candidate NGC\,4736\_2 shows the spectral lines with the P Cyg profile and some spectral and photometric variability ($\Delta$V $\approx 0.5$ mag, $\Delta$B $\approx 0.4$  mag). Such a small variation does not allow us to classify NGC\,4736\_2, so it remains in the status of the LBV candidate. 

The object NGC\,4736\_1 shows a significant brightness variation, the difference between the extrema of the light curve were $\Delta$V $\approx 1.1$ mag and $\Delta$B $\approx 0.82$ mag in both bands. Based on its spectrum typical of the LBV, the characteristic bolometric luminosity and considerable brightness variations, we conclude that NGC\,4736\_1 is actually  
the LBV star. More accurate estimating of the temperature of the stars, their bolometric luminosities and searching for their variability, require further observations and  more detailed analysis of the available data. 
\section*{Acknowledgments} 

This research was supported by the Russian Scientific Foundation (N 14-50-00043) as for observations and was partly supported by the Russian Foundation for Basic Research (N 18-32-00333, 16-02-00758, 16-02-0567) in part of data analysis. The authors acknowledge partial support from M.V.Lomonosov Moscow State University Program of Development.

\bibliographystyle{mnras} \bibliography{bibtexbase.bib} 

% if your bibtex file is called example.bib
%\begin{thebibliography}{99}
%\end{thebibliography}
%%%%%%%%%%%%%%%%%%%%%%%%%%%%%%%%%%%%%%%%%%%%%%%%%%
%%%%%%%%%%%%%%%%% APPENDICES %%%%%%%%%%%%%%%%%%%%%
%\appendix
% Don't change these lines
\bsp	\label{lastpage} \end{document}